\pdfoutput=1

\documentclass[aps,prl,reprint,showpacs,superscriptaddress]{revtex4-1}

\usepackage{graphicx}
\usepackage{graphics}
\usepackage{amsmath}
\usepackage{amssymb}
\usepackage{amsfonts}
\usepackage{dcolumn}
\usepackage{dsfont}
\usepackage{latexsym}
\usepackage{rotating}
\usepackage{color}
\usepackage{latexsym}
\usepackage{bbm}
\usepackage{subfigure}
\usepackage{float}
\usepackage{epsfig}
\usepackage{epsf}
\usepackage{psfrag}
\usepackage{bm}
\usepackage{amsthm}
\usepackage{eucal}
\usepackage{mathrsfs}
\usepackage{url}
\usepackage{braket}
\usepackage{epstopdf}
\graphicspath{ {./images/} }
\newcommand{\RNum}[1]{\uppercase\expandafter{\romannumeral #1\relax}}
\usepackage{color} 

\usepackage{hyperref}
\hypersetup{
colorlinks=true,final=true,
        linkcolor=blue,
        citecolor=blue,
        filecolor=blue,
        urlcolor=blue,
}
  
\begin{document}

\title{Fully spin-polarized two dimensional polar semi-metallic phase in Eu substituted GdI$_2$ monolayer}
  
\author{Srishti Bhardwaj}
\affiliation{ Department of Physics, Indian Institute of Technology Roorkee, Roorkee - 247667, Uttarakhand, India}
\author{T. Maitra}
\email{tulika.maitra@ph.iitr.ac.in}
\affiliation{ Department of Physics, Indian Institute of Technology Roorkee, Roorkee - 247667, Uttarakhand, India}

\date{\today}

\begin{abstract}
The coexistence of seemingly mutually exclusive properties such as ferromagnetism, ferroelectricity and metallicity in atomically thin materials is the requirement of the hour in electronics as the Moore's law faces an impending end. Only a few 2D multiferroic materials have been predicted/realized so far. The polar metals with simultaneous presence of polarity and conductivity are also equally rare. Here, we predict, based on first-principles calculations that an Eu-substituted rare-earth halide GdI$_2$ monolayer showcases ferromagnetism, ferroelasticity while being polar and a fully spin-polarized semi-metal at the same time. The ferroelasticity and polarity are shown to be coupled making it possible to switch the polar direction using external mechanical stress. Further, it is observed that, an application of biaxial tensile strain of $5\%$ causes the spin easy-axis to shift from out-of-plane to in-plane direction. Thus spin easy axis gets coupled with the direction of polarization in the strained monolayer making the switching of magnetization also possible using external strain. Simultaneous coexistence and coupling of the ferroic orders in a metallic 2D material makes the Eu substituted GdI$_2$ monolayer an incredibly rare material for nano-electronics and spintronics applications.

\end{abstract}

\maketitle

Two-dimensional multiferroics have drawn huge attention lately due to their potential applications in nano-electronics and spintronics even though very few have been realized so far due to formidable challenges\cite{Ramesh, Tang, An}. Ferroic orders such as ferromagnetism and ferroelectricity are rarely found in one system simultaneously because of their mechanisms being mutually exclusive\cite{Hill}. Similarly, polar metals with simultaneous presence of ferroelectricity (or polarity) and metallic conductivity (two contra-indicative properties) are very rare\cite{W. Zhou}. The itinerant electrons present in a metal are expected to screen the long-range Coulomb interactions responsible for dipole alignment. Therefore, a simultaneous existence of polarity, ferromagnetism and metallic conductivity is indeed a very rare occurence while being very important from the novel physics perspective. Also, such materials hold great potential for device applications.

In 1965, the possibility of coexistence of metallicity and ferroelectricity/polarity in a material was first given by Anderson and Blount based on the condition that the itinerant electrons do not interact strongly with the transverse optical (TO) phonons responsible for ferroelectricity\cite{Anderson}. The search for such materials was recently revitalized by the prediction of a polar metal LiOsO$_3$ in 2013 by Shi et. al.\cite{Shi} and later on by the experimental discovery of electric-field switchable ferroelectric metal WTe$_2$\cite{Fei, Sharma}. The coexistence of metallicity and polarity in these materials was explained by the `decoupled electron mechanism' proposed by Puggioni and Rondinelli\cite{Puggioni} which is an extension of the original Anderson-Blount model of `weak coupling between electrons at Fermi level and soft TO phonons'\cite{Anderson}.

In another recent development, rare-earth Gd based halide monolayers (GdX$_2$ (X=Cl, Br, I)) have been predicted theoretically\cite{B. Wang,Liu} which possess large magnetic moments of about 8$\mu_B$ per formula unit (f.u.) and predicted to have large Curie temperatures (more than 220K). Subsequently, a coexistence of two ferroic orders : ferroelasticity and ferromagnetism was predicted, from first principle calculations, by You et al.\cite{You} via electron doping in GdI$_3$ monolayer. In our previous work\cite{SB1}, we predicted the coexistence of three ferroic orders: ferroelasticity, ferroelectricity and ferromagnetism in hole doped GdCl$_2$ monolayer via Eu-substitution. However, in all these previous predictions of multiferroic materials (with more than one ferroic order), the materials were insulating in nature.   

In this work, we show that a similar Eu-substitution in GdI$_2$ monolayer results in a ferroelastic (FA), structurally polar 2D material which is semi-metallic with both electrons and holes participating in the conduction. The counter-intuitive coexistence of conduction and polarity can be explained by a charge ordering phenomenon which involves a lowering in electronic energy of the material and doesn't require long range Coulomb forces for the dipole alignment. It is also shown that an application of a biaxial tensile strain of $5\%$ on the ferromagnetic (FM) monolayer shifts the spin easy-axis from out-of-plane of the monolayer (PMA) to in-plane(IMA). Due to the directional dependence of magnetic anisotropy in the monolayer plane, the spins are preferably aligned in a direction perpendicular to the polar axis. This results in a coupling of the ferromagnetism and ferroelasticity/polarity of the structure, highly desired for multifunctional device applications\cite{Palneedi, Eerenstein}.

We performed the first-principles calculations using spin-polarized density functional theory (DFT) as implemented in the Vienna \textit{ab-initio} Simulation Package(VASP)\cite{Kresse}. The core electrons are dealt with using the projector augmented wave (PAW) method. We have used the Perdew-Burke-Ernzerhof (PBE) parametrization\cite{Perdew} of the generalized-gradient approximation as the exchange correlation functional. The energy cut-off for the plane wave basis was 500 eV. To account for the Hubbard correlation in Gd and Eu 4f-bands, we used the rotationally invariant GGA (PBE) + U method \cite{Liechtenstein} with U$_{eff}$ = U - J = 8.0 eV (here U is Coulomb correlation and J is Hund's exchange) for both Gd and Eu 4f-orbitals, the value taken from previous work\cite{B. Wang}. We have varied the U$_{eff}$ value in the range 4 - 10 eV and observed that our results presented below do not change qualitatively. Treatment of f-electrons using PBE+U is possible as f-subshell is half filled. We inserted a vacuum of 18.5 $\AA$ between two layers to avoid interlayer interactions. The Hellmann-Feynman forces convergence criterion is set at 0.01 eV/$\AA$ while relaxing the structure and the energy convergence criteria is $10^{-6}$ eV. A $\Gamma$ centered $11\times11\times1$ Monkhorst-Pack grid was chosen for structural optimization. The phonon dispersion calculations were performed using the Density Functional Perturbation Theory (DFPT) as implemented in VASP and the PHONOPY package with a $3\times3\times1$ supercell\cite{Gonze,Togo}. To check thermal stability, \textit{ab-initio} molecular dynamics (AIMD) calculations were performed where we use NPT ensemble using Langevin thermostat at zero pressure and at 50K\cite{Hoover, Evans}. The simulation was performed for 12 ps with a time step of 3 fs. The electrical conductivity is calculated by solving the semiclassical Boltzmann transport equations based on constant relaxation-time approximation (here $\tau =10 fs$) as implemented in the BoltzWann code \cite{Pizzi}. We used a $7\times7\times3$ {\bf k}-points grid to construct the Maximally Localized Wannier Functions (MLWFs), while a denser $120\times120\times120$ k-points grid was used to calculate the electrical conductivity. 
\begin{figure}[!t]
\centering
\includegraphics[width=9cm]{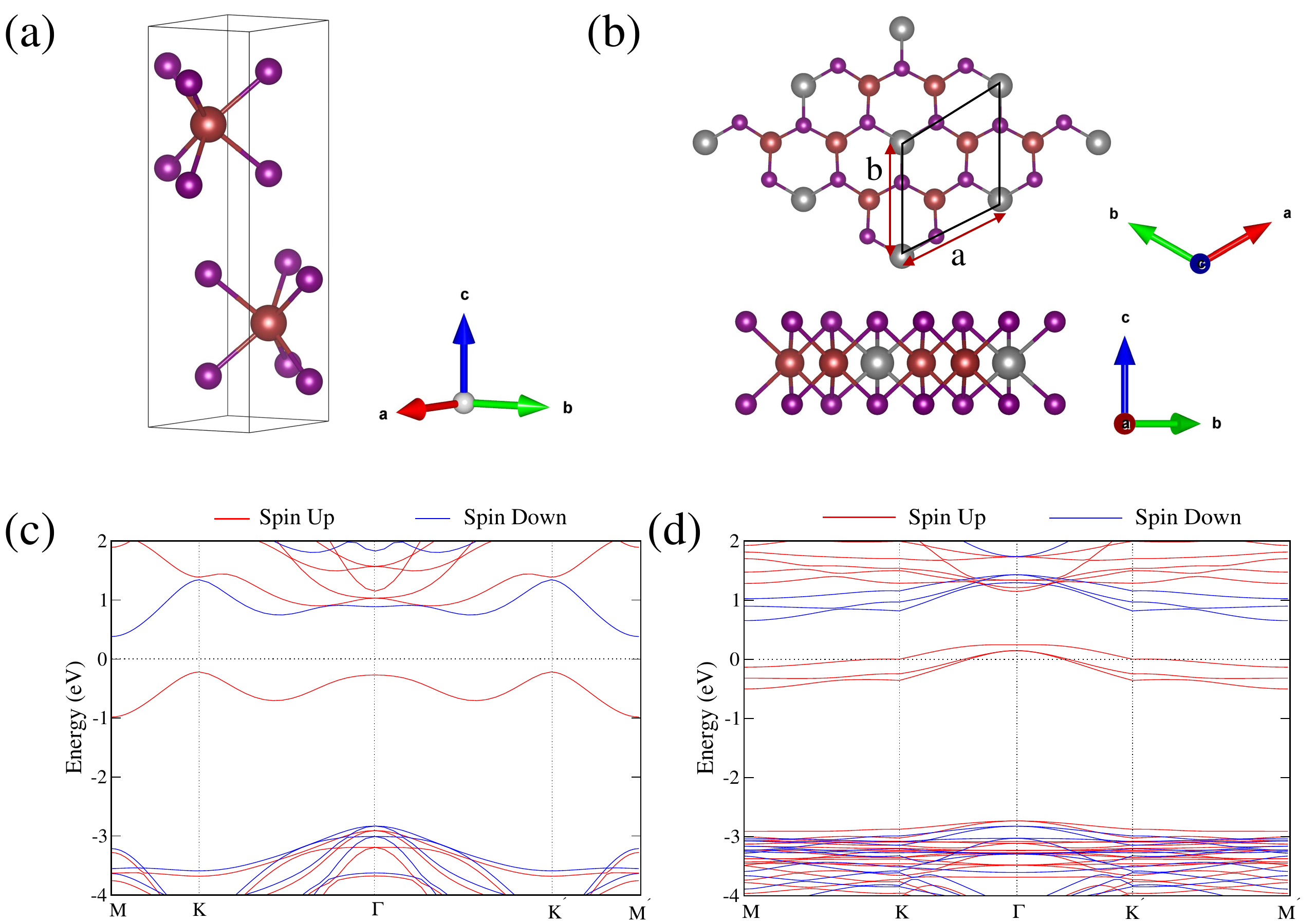}
\caption{(a) Unit cell of bulk GdI$_2$ with 2H-MoS$_2$ structure (brown spheres represent Gd atoms whereas violet spheres represent I atoms). (b) Top and side views of high symmetry hexagonal (P$\overline{6}$m2) Eu-substituted GdI$_2$ monolayer (grey spheres denote Eu atoms). (c) Spin-resolved electronic band structure of pristine GdI$_2$ monolayer. (d) Spin-resolved electronic band structure of hexagonal Eu-substituted GdI$_2$ monolayer.}
\end{figure}

Bulk GdI$_2$ is ferromagnetic with a high Curie temperature of about 290K and a large saturation magnetic moment of about 7.33 $\mu_B$ per Gd atom near room temperature\cite{Felser, Ahn}. It has a vdW layered 2H-MoS$_2$ type structure, in which the I-Gd-I sandwich layers are stacked together. It is predicted to be easily exfoliable to create a monolayer. We calculated the relaxed equilibrium lattice constants of monolayer (ML) GdI$_2$ which are found to be a = b = 4.099 Å, quite close to the previously reported values \cite{Cheng}. The Gd atoms form a 2D triangular lattice and each Gd is trigonal prismatic coordinated with six I atoms. From our calculations, we found it to be a FM semiconductor with a band gap of 1.12 eV, the findings agreeing with the earlier reports\cite{B. Wang}.  We present the spin-resolved electronic band structure of GdI$_2$ monolayer in Fig.1(c) which clearly shows the semiconducting nature. The oxidation state of Gd ions in ML GdI$_2$ is found to be +2 resulting in the electronic configuration to be $4f^75d^1$.

We attempted a hole doping in the GdI$_2$ monolayer by substituting the central Gd atoms in all the Gd hexatomic rings by Eu atoms (see Fig. 1(b)). The Eu ions having the same oxidation state as Gd ions (i.e +2) have an electronic configuration $4f^75d^0$ and amount to a 1/3rd hole doping in the monolayer. The band structure presented in Fig. 1(d) clearly shows the presence of hole doped spin up bands around the Fermi level making the Eu-substituted GdI$_2$ monolayer metallic in nature. In order to ensure the feasibility of this substitution, we calculated the formation energy of the final structure following the method used in our previous work\cite{SB1}. The value of formation energy was found to be -4.75 eV/f.u. under I-rich condition and -0.41 eV/f.u. under Gd-rich condition which indicates the substitution process is energetically favourable.

In order to check its dynamic stability, we calculated its phonon spectrum and observed soft phonons (see Fig. 2(a)) which indicates that the high symmetry (HS) hexagonal structure (P$\overline{6}$m2) is unstable and would spontaneously transition to some other lower symmetry (LS) state. To find this stable LS state, we employed the atomic displacement eigenvectors corresponding to the soft phonon modes shown in Fig. 2(b) into the hexagonal structure and performed further optimization. The resulting LS structure was found to have the polar Amm2 space group where it displayed expanding and contracting of Gd-Gd `bonds' causing an elongation in one direction of the structure (lattice parameter $a$) and a simultaneous contraction in its perpendicular direction (lattice parameter $b$) (Fig. 3(a)). Also, it shows a shifting of the Eu atoms away from their initial positions i.e. the center of the Gd hexagonal ring, thus breaking the centrosymmetry of the structure. The optimized lattice parameters of the final structure alongwith those of initial HS structure are given in Table I. When compared with the original high symmetry hexagonal structure, it is found to have a spontaneous strain \(((a-b)/a\times100)\) of about 1\%.  The phonon spectrum for the LS Amm2 structure isn't found to have any soft phonons indicating its dynamic stability (Fig. S1(a)).

\begin{figure}[!b]
\centering
 \includegraphics[width=9.0cm]{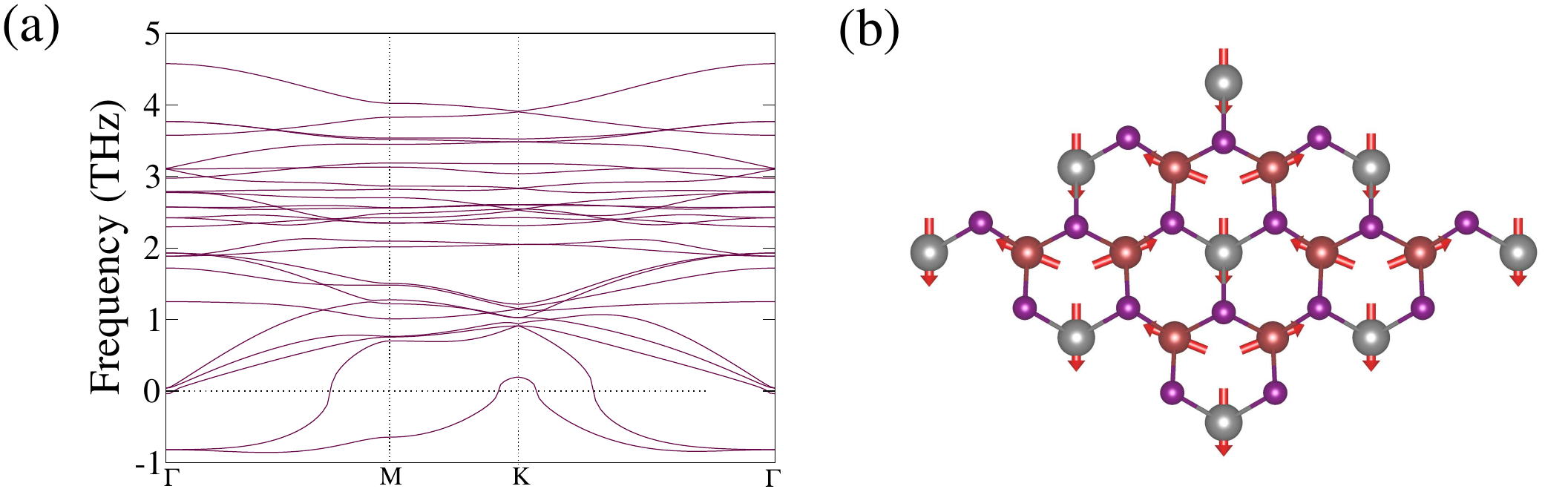}
 \caption{(a) Phonon spectrum of high symmetry hexagonal Eu-substituted monolayer with imaginary frequency (soft) phonon modes. (b) Atomic displacement eigenvectors corresponding to soft phonon modes.}
\end{figure}


\begin{table}[!h]
\centering
\renewcommand{\arraystretch}{1.5}
\label{tab1}
\caption{Lattice parameters of HS hexagonal (P$\overline{6}$m2), LS Amm2 and monolayer under 5\% biaxial tensile strain.}

\scalebox{1.1}{
\begin{tabular}{|c|c|c|c|c|c|c|}
\hline
\textbf{Structure} & \textbf{$d_1(\AA)$} & \textbf{$d_2(\AA)$} & \textbf{$d_3(\AA)$} & \textbf{$d_4(\AA)$} & \textbf{$a(\AA)$} & \textbf{$b(\AA)$} \\
\hline
Hexagonal & 4.29 & 4.29 & 4.29 & 4.29 & 7.44 & 7.44 \\
\hline
Amm2 & 4.65 & 4.14 & 4.56 & 4.18 & 7.48 & 7.41 \\
\hline
Strained & 4.90 & 4.34 & 4.83 & 4.38 & 7.85 & 7.79 \\
\hline
\end{tabular}}
\end{table}
\begin{figure*}[!t]
\centering
 \includegraphics[width=18.0cm]{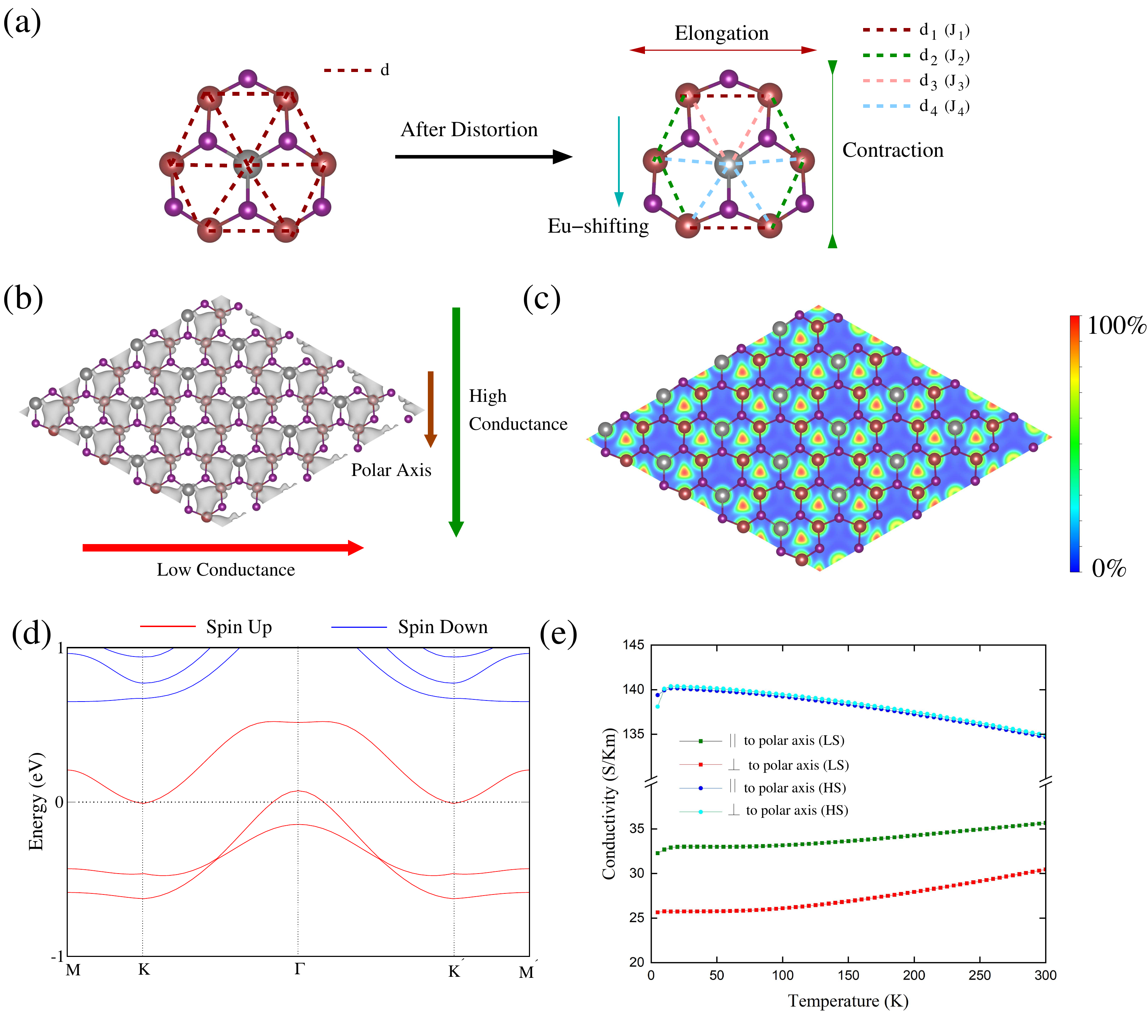}
 \caption{(a) Gd hexagonal ring with Eu at the center in HS structure distorting to the LS structure due to BCO (Top view). Here brown and grey spheres represent Gd and Eu atoms respectively. (b) Partial charge density distribution corresponding to states near the Fermi level showcasing bond-centered charge ordering (BCO). (c) Electron localization function along the monolayer plane consisting of Gd and Eu-ions. (d) Spin-resolved electronic band structure of low symmetry (space group Amm2) Eu-substituted monolayer (e) Electrical conductivity vs Temperature of HS and LS structures along and perpendicular to the polar axis of the LS Amm2 monolayer.}
 
\end{figure*}

In order to explain the microscopic mechanism involved behind the Gd-Gd bond-length change, we looked at the partial charge density distribution corresponding to the energy states close to the Fermi level, of the LS Amm2 structure (Fig.3(b)). We see that there is a bond centered charge ordering (BCO) in the final LS structure. The bonds with finite charge density between them become shorter whereas the others become longer. In Fig. 3(c) we present the electron localization function (ELF) which clearly shows charge density localization at the center of the bond instead of at the atomic sites, further establishing BCO in this system. However, we observe that despite there being a BCO, the final LS structure remains a half (fully spin-polarized) semi-metal as is indicated by the spin-up electron and hole pockets in the electronic band diagram (Fig. 3(d)). The semi-metallic nature remains robust up to U$_{eff}$ = 10 eV. Thus, the Eu-substituted GdI$_2$ ML can be categorised as a polar metal. This rare counter-intuitive polar character in the metallic ML is most probably governed by the lattice distortion associated with BCO. This distortion not only involves the shortening of Gd-Gd bonds where the charge density is predominantly localized but also causes a simultaneous off-centre shifting of Eu-ions. Thus, the polar Eu shifting doesn't require any long range Coulomb interactions as is usually the case in conventional ferroelectrics and the polarity in the structure is caused by a chemical bonding phenomenon which lowers the electronic energy of the material. This fact makes it possible for the structure to become polar despite it being a metal. We would like to note that Eu-substituted GdCl$_2$, on the other hand, becomes insulating in nature\cite{SB1}.  

In order to check the screening of charge on Eu ions by the conduction electrons, we calculated the Born effective charge (BEC) of the Eu ions for the HS hexagonal structure and the LS Amm2 structure using density functional perturbation theory as implemented in VASP\cite{Gajdos}. It is clear that the BEC of Eu$^{2+}$ ions in the hexagonal structure is very small (see Table II) implying that the charge on Eu ions is almost completely compensated by the conduction electrons in the hexagonal structure. However, in the LS structure, the BEC of Eu ions is quite high indicating only a partial screening of the charge. This partial compensation of charge on Eu$^{2+}$ ions further suggests that the polar ML may be able to respond to an external electric field despite the presence of conduction electrons. 

This partial screening can also be explained by the lower conductance values of the LS structure (being a semi-metal) as compared to that of the HS one as can be seen in Fig. 3(e). Also, it is clear from the partial charge density distribution of LS structure (Fig. 3(b)) that the conductance in the LS structure has a slightly anisotropic nature which can also be seen in Fig. 3(e). The conductance along the polar axis is more whereas perpendicular to the polar axis, it is less. This anisotropy of conductance is also supported by the BEC value of Eu ions which is smaller along the polar axis and larger perpendicular to it (see Table II).

\begin{table}[!h]
\centering
\renewcommand{\arraystretch}{1.5}
\label{tab1}
\caption{Born Effective Charge of Eu ions along and perpendicular to the polar axis in units of $e$ i.e. electronic charge.}

\scalebox{1.0}{
\begin{tabular}{|c|c|c|}
\hline
\textbf{Structure} & \textbf{HS (P$\overline{6}$m2)} & \textbf{LS (Amm2)} \\
\hline
Along the polar axis & 0.54 & 1.56  \\
\hline
Perpendicular to the polar axis & 0.54 & 2.34  \\
\hline
\end{tabular}}
\end{table}
\begin{figure}[!ht]
\centering
 \includegraphics[width=8.5cm]{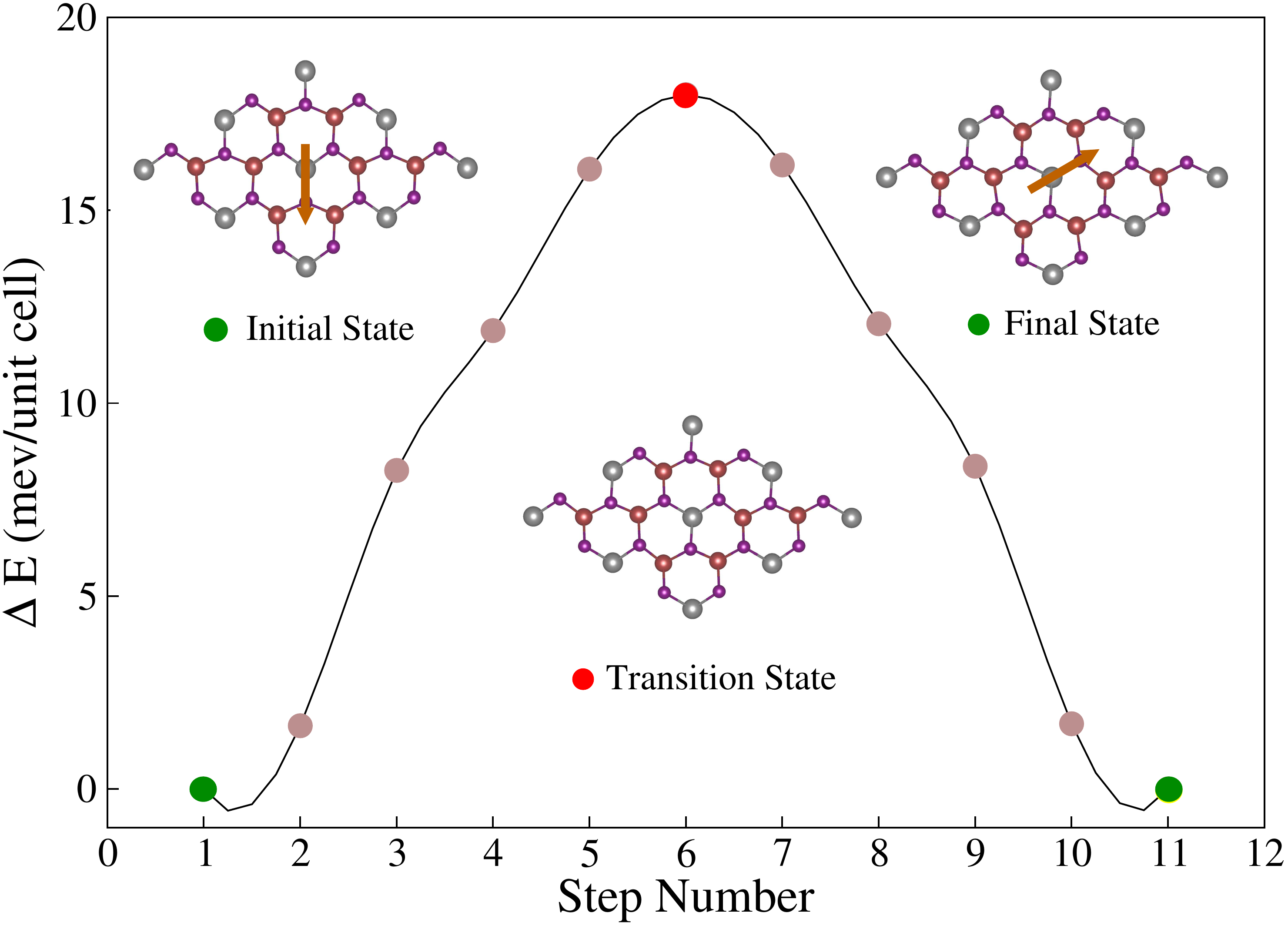}
 \caption{Minimum energy pathway for transition from one ferroelastic(polar) phase to another obtained using solid state nudged elastic band (SS-NEB) method. In the insets are initial, final and highest energy transition states (Brown arrow in initial and final states represents the polar axis).}
\end{figure}

 The HS structure being three-fold rotationally symmetric can have a lattice distortion along three possible directions. Therefore, the strain direction should be switchable by the application of an external stress, thus making the structure ferroelastic. Moreover, the polar axis of the ML is directly coupled to the direction of the strain and therefore, should switch along with the strain direction. We found, by first-principle calculations, that a tensile strain of about 7 $\%$ along one of the shorter Gd-Gd bonds and an equal simultaneous compressive strain in the perpendicular direction results in a phase transformation with the polar axis direction rotated by 120$^\circ$. Thus, the Eu-substituted GdI$_2$ ML can be termed as a ferroelastic-polar metal in which the polar axis can be rotated by an external mechanical stress. In order to investigate the energetics of the structural transition from one ferroelastic(polar) state to another, we performed a solid-state nudged elastic band (SS-NEB) calculation\cite{Sheppard}. The minimum energy pathway (MEP) for transition from one lattice configuration to another rotated at 120$^\circ$ is shown in Fig. 4. The transition barrier height is 18 meV/formula unit (Gd$_2$EuI$_6$) which is of the same order as in 2D FE-FA multiferroic MX monolayers (M=Sn, Ge; X=S, Se)\cite{H. Wang}. Such small energy barrier points toward an easy and fast switching between different phases.

We expect the ground state magnetic configuration of the monolayer to be ferromagnetic (FM) owing to its metallic character. To confirm this, we studied four different magnetic configurations along with the FM one (Fig. S3) using GGA+U+SO. Spin-orbit coupling was considered as we are dealing with heavy rare-earth ions. The energies of the magnetic configurations \RNum{2}, \RNum{3}, \RNum{4}, \RNum{5} are found to be 662.4 meV, 147.3 meV, 647.6 meV and 267.4 meV per magnetic unit cell (shown in Fig. S3) higher than that of FM (\RNum{1} in Fig. S3), establishing FM as the ground state magnetic ordering. The value of magnetization per formula unit is found to be 21.58$\mu_B$ which is consistent with the formal magnetic moment of 8$\mu_B$ on Gd with electronic configuration \(4f^7d^1\) and 7$\mu_B$ on Eu with electronic configuration \(4f^7d^0\), in valence state +2. The spin easy-axis is found to be in the out-of-plane direction. We calculated the magnetic anisotropy energy using the formula $MAE = E_\theta-E_c$, where $\theta$ is the polar angle in the plane of the monolayer (ab plane) measured with respect to polar axis (see Fig. 5(a)). Interestingly, it is observed that the MAE is not the same in all the directions in-plane of the ML as is usually the case with centrosymmetric MLs. Instead, we find that the MAE value depends upon the direction of polar axis (Fig. 5(b)). Even though the spin easy-axis remains out of plane, the preferred direction inside the plane of the monolayer is the one perpendicular to the polar axis. The magnetic anisotropy energy (MAE) corresponding to the direction perpendicular to polar axis ($\theta=90^\circ$) is found to be about 806 $ \mu $eV/f.u. which is pretty high when compared with that of pristine GdX$_2$ monolayers\cite{Liu} and hence can play a significant role in stabilizing the magnetic ordering in the 2D limit. The values of exchange parameters of the Eu substituted monolayer are found to be higher than those of the pristine GdI$_2$ monolayer (see Supplemental Material). Thus, one can expect a fairly large ferromagnetic transition temperature (T$_c$).
\begin{figure*}[!t]
\centering
\includegraphics[width=19cm]{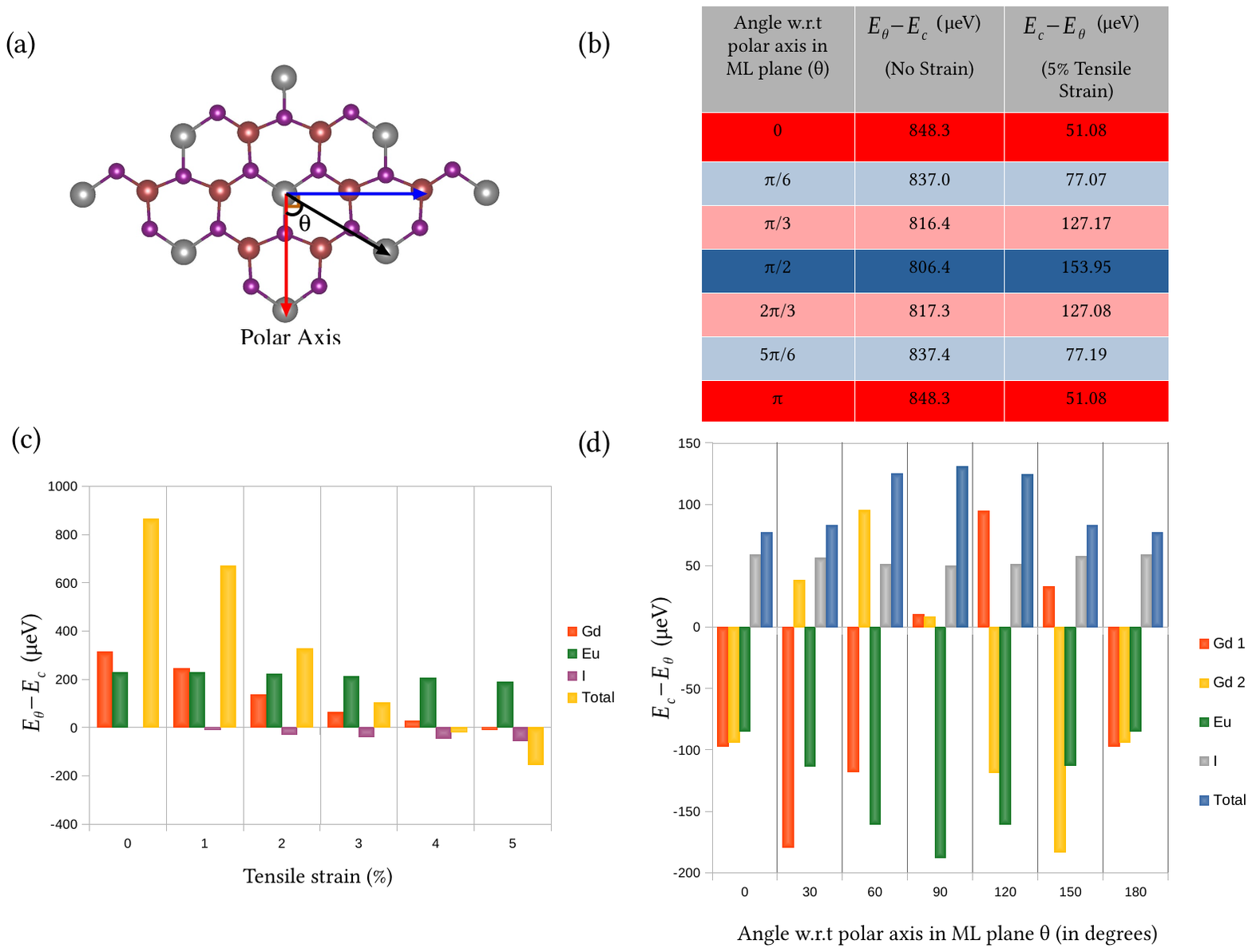}
\caption{(a) Different directions inside the plane of monolayer can be described in terms of the angle (\(\theta\)) they make with the polar axis. (Red arrow denotes the polar axis direction, blue arrow the direction perpendicular to the polar axis and the black arrow denotes any random direction in the monolayer plane.) (b) Magnetic anisotropy energy (MAE) along different directions in the monolayer plane without strain and with 5\% biaxial tensile strain (MAE is defined as \(E_\theta-E_c\) without strain and as \(E_c-E_\theta\) with strain.) (c)  Variation of magnetic anisotropy energy defined by \(E_\theta-E_c\) with biaxial tensile strain. (d) Atomic contribution to MAE along different directions in ML plane under 5\% tensile strain.} 
\end{figure*}

Now, as the preferred spin orientation is out-of-plane of the monolayer, there is no coupling between magnetization and the other order parameters i.e. strain and polarity. However, it has been observed in case of GdX$_2$ MLs\cite{You} that the spin-easy axis shifts from being out-of-plane to in-plane on the application of a tensile strain. Expecting similar trend, we applied a biaxial tensile strain on the monolayer. We observed that, on increasing the value of strain, the MAE decreases and eventually becomes negative at a value of $5\%$ i.e. the spin-easy axis turns in-plane(Fig. 5(c)). The lattice parameters of the monolayer under 5\% biaxial tensile strain are given in Table I (last row). As previously mentioned, the MAE is anisotropic inside the plane of the polar ML and is found to be maximum in a direction perpendicular to the polar axis ($\theta=90^\circ$). Hence, the spin-easy axis turns out to be along $\theta=90^\circ$ in the monolayer plane (Fig. 5(b)). Thus, as a consequence of the polar nature of the ML structure, the spin easy-axis of the monolayer is coupled with its polar axis which in turn is linked with the strain direction. Thus, under  $5\%$ tensile strain, the ferroelasticity, polarity and ferromagnetism of the ML become coupled and a change in the strain direction by external mechanical stress results in a change in the direction of polar axis and hence in that of magnetization. However, it should be noted that the substituted ML becomes a half-semiconductor at $5\%$ tensile strain as shown in the band structure (Fig. S1(d)), although it retains all the previously mentioned ferroic orders. We calculated the spontaneous polarization value of the ML under 5\% tensile strain following similar method as in our previous work \cite{SB1} and found it to be 43.72  pC/m along Eu-shifting direction. The phonon spectrum of the strained ML is shown in Fig. S1(c) which shows no soft phonons confirming its dynamic stability.

The microscopic origin of the MAE change under strain can be explained by element-resolved MAE. The MAE contribution corresponding to a given orbital of an atom can be obtained from the spin-orbit matrix element difference under the second order perturbation theory\cite{X. Wang}:
\[MAE= \zeta^2 \sum_{o,u} \frac{|\psi_o|\hat{L}_x|\psi_u|^2- |\psi_0|\hat{L}_z|\psi_u|^2}{E_o-E_u}\]
where $\psi_o$ and $\psi_u$ are the occupied and unoccupied orbitals respectively, $E_o$ and $E_u$ being the corresponding energies and $\zeta$ is the spin-orbit coupling constant. We calculate the atom-resolved MAE by summing the contribution from all the valence orbitals of the atom. As shown in Fig. 5(c), it can be seen that the MAE contribution from Gd atoms favors perpendicular magnetic anisotropy (PMA) in the absence of any strain which, in the presence of strain, decreases monotonically and eventually becomes negative, thus, favouring in-plane magnetic anisotropy (IMA) instead. On the other hand, the I-atom MAE contribution is a small positive value (favouring PMA) without strain which quickly becomes negative and increases rapidly in the negative direction on increasing the tensile strain. The Eu-atom MAE also decreases, although slowly, in favour of IMA upon the application of tensile strain. Thus, a combined effect of these MAE contributions ultimately favours an in-plane spin easy-axis at a biaxial tensile strain of $5\%$.

In order to explain the anisotropic MAE inside the plane of the monolayer, we need to compare the specific atom MAE contributions. As shown in Fig. 5(d), the MAE contribution from both Gd ions favours IMA only along the direction perpendicular to the polar axis ($\theta=90^\circ$). In all the other directions, the MAE of at least one of the two Gd atoms favours PMA instead. This behavior from Gd ions results in the total MAE to be the highest for the $\theta=90^\circ$ direction making it most favourable for spin orientation. The above mentioned trend of MAE contribution from Gd ions follows from the anisotropic charge density distribution (Fig. (3(b)) caused by the bond-centered charge ordering (BCO) in the ML.

In summary, we have performed a detailed investigation using first principles DFT calculations on Eu substituted GdI$_2$ monolayer. The Eu substituted monolayer undergoes a bond-centered charge ordering making it transition to a lower symmetry (space group Amm2) distorted structure with long and short Gd-Gd bonds along with a shifting of Eu-ions from their centrosymmetric position. This gives rise to remarkable new physical phenomena such as ferroelasticity and polarity in the system which are found to be switchable via external mechanical stress. Most interestingly, the electronic structure reveals a fully spin-polarized semi-metallic nature with both electrons and holes as charge carriers which makes the system a 2D polar/ferroelectric-like metal while being ferromagnetic and ferroelastic, an incredibly rare coexistence of multiferroic orders and metallicity. Further, we have observed that with the application of 5$\%$ tensile strain the out-of-plane spin easy axis turns to in-plane and hence gets coupled to the polar axis. The strained monolayer however becomes insulating in nature. The coexistence of coupled ferroic orders in the strained monolayer can be immensely useful for device applications.     

\noindent SB acknowledges Council of Scientific and Industrial Research (CSIR), India for research fellowship. TM acknowledges Science and Engineering Research Board (SERB), India for funding support through MATRICS research grant (MTR/2020/000419). TM and SB acknowledge PARAM GANGA computational facility at IIT Roorkee.

\end{document}


\title{            Supplemental Materials to\\  Fully spin-polarized two dimensional polar semi-metallic phase in Eu substituted GdI$_2$ monolayer}
\author{Srishti Bhardwaj}
\affiliation{ Department of Physics, Indian Institute of Technology Roorkee, Roorkee - 247667, Uttarakhand, India}
\author{T. Maitra}
\email{tulika.maitra@ph.iitr.ac.in}
\affiliation{ Department of Physics, Indian Institute of Technology Roorkee, Roorkee - 247667, Uttarakhand, India}

\date{\today}
\maketitle

\setcounter{figure}{0}
\renewcommand{\figurename}{Fig.}
\renewcommand{\thefigure}{S\arabic{figure}}
\setcounter{table}{0}
\renewcommand{\tablename}{Table}
\renewcommand{\thetable}{S\arabic{table}}

The phonon dispersion spectra for both un-strained and strained (5\% tensile strain) monolayer (ML) show no soft phonon modes ensuring the dynamic stability of the Amm2 monolayer structure with and without strain (see Fig. S1 (a) and (c)). The projected density of states of the monolayer (Fig. S1 (b)) shows half-metallic nature with the states close to the Fermi level mainly comprise of Gd $d$ orbitals with a small hybridization with Eu $d$ and I $p$ orbitals. The strained monolayer becomes insulating as can be seen from the band structure presented in Fig. S1 (d). 
\begin{figure}[!h]
\centering
 \includegraphics[width=18.0cm]{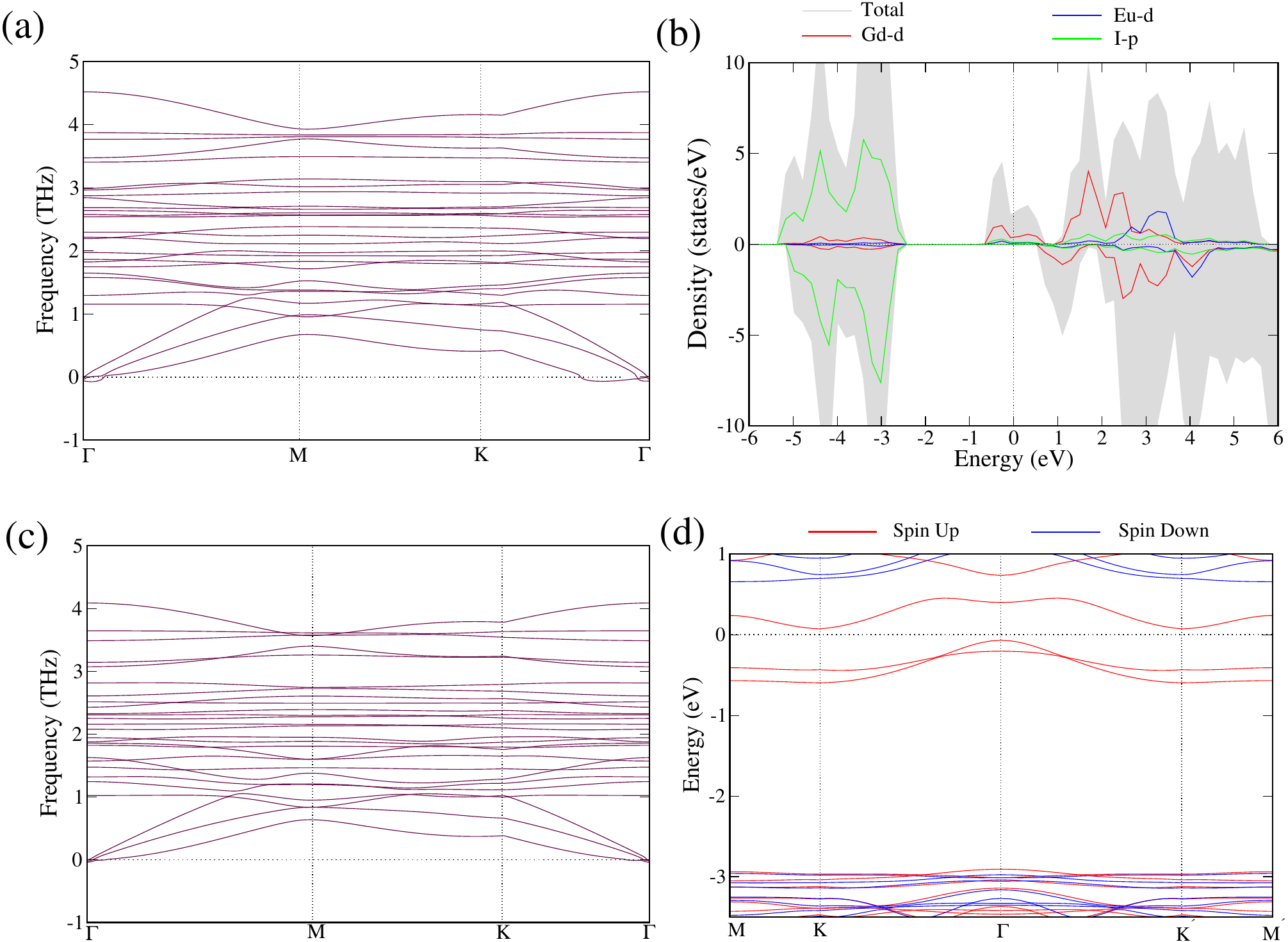}
 \caption{(a) Phonon spectrum of Amm2 structure of Eu-substituted GdI$_2$ monolayer. (b) Projected density of states (DOS) of the monolayer. (c)Phonon spectrum and (d) spin-resolved electronic band diagram of ML under $5\%$ biaxial tensile strain.}
\end{figure}

\section{AIMD simulation for un-strained and strained doped monolayers at 50K}
In order to check thermal stability of the Eu-substituted GdI$_2$ monolayer, we performed the \textit{ab-initio} molecular dynamics (AIMD) simulation at a temperature of 50 K for 12 ps with a time step of 1 fs. The energy fluctuations are small and the structure remains almost intact after 12 ps [Fig. S2(a)]. We performed the similar calculation for monolayer under 5\% biaxial tensile strain also and arrived at a similar result [Fig. S2(b)].
\begin{figure}[!h]
\centering
 \includegraphics[width=18.0cm]{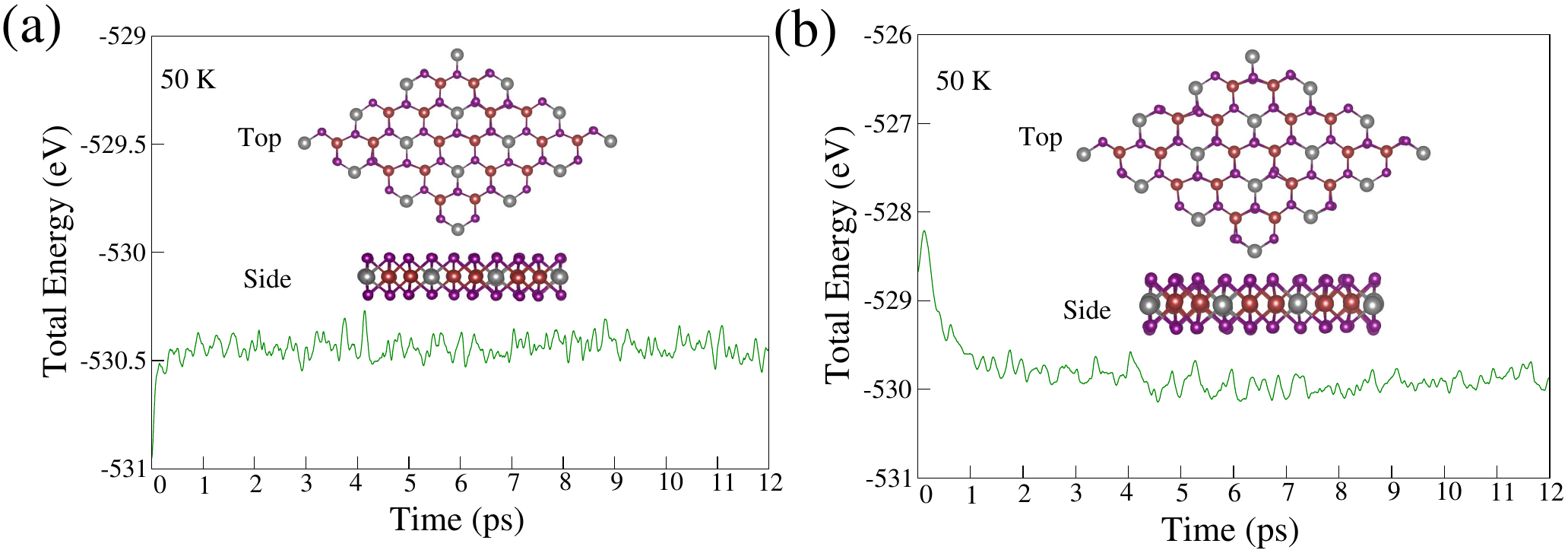}
 \caption{Energy fluctuations and structure snapshots after 12 ps at 50K for (a) un-strained and (b) 5\% tensile strained ML.}
\end{figure}

\section{Calculation of Magnetic Exchange Parameters}
\begin{figure}[!h]
\centering
 \includegraphics[width=10.0cm]{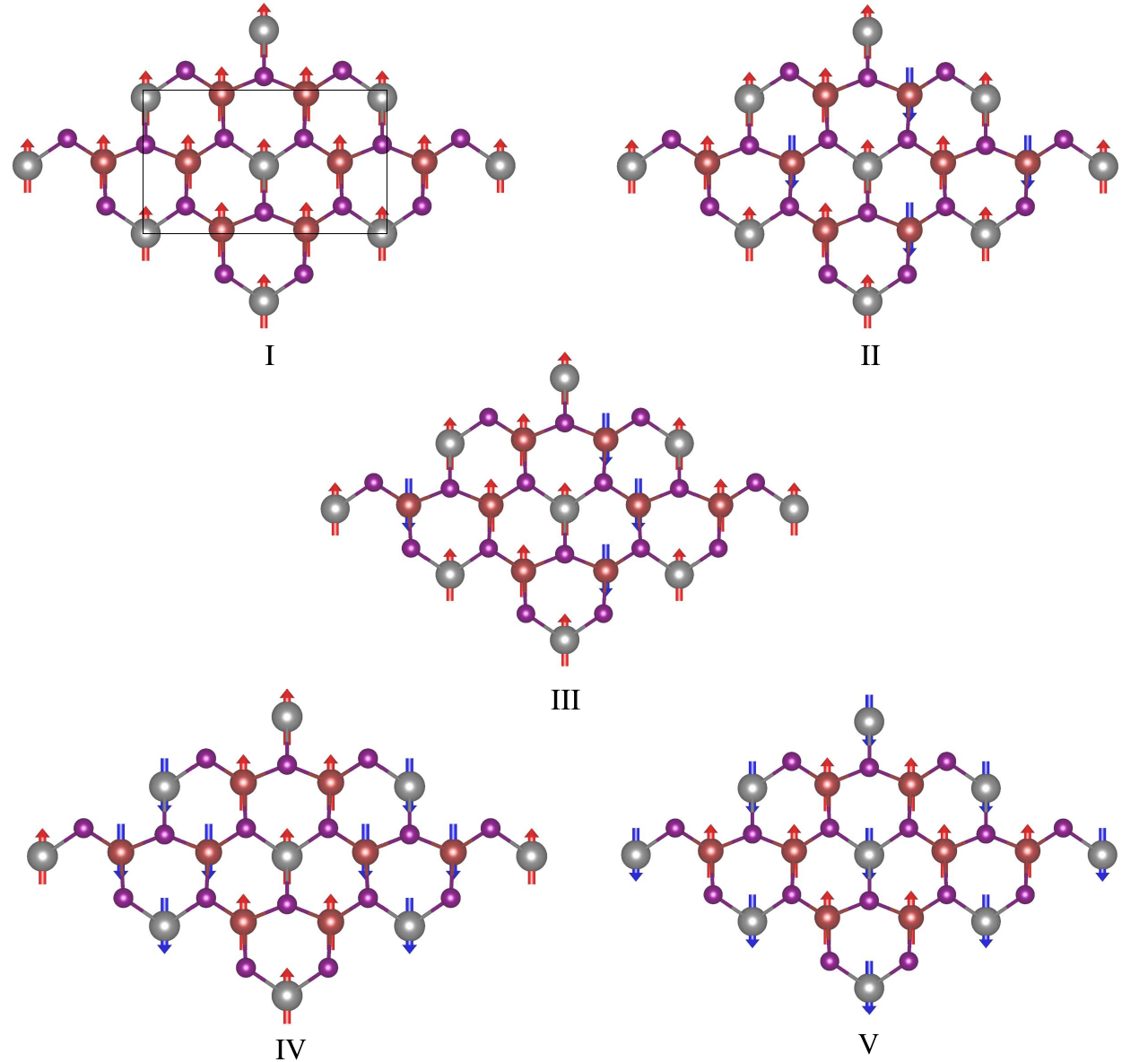}
 \caption{Magnetic configurations used to calculate exchange parameters (black rectangle represents magnetic unit cell)}
\end{figure}

\vspace{0.5cm}	

We have calculated the magnetic exchange interaction parameters following the method used in our earlier work\cite{SB1}. 
Because of the bond centered charge ordering, the Gd-Gd bonds with finite charge density between them are shorter and the others are longer. Similarly, due to Eu off-center shifting, all the Eu-Gd bonds are also not equal in length as shown in Fig. 3(a) in the main manuscript. Thus, in the Amm2 structure of the monolayer, there are four kinds of NN exchange parameters denoted by \(J_1\), \(J_2\), \(J_3\) and  \(J_4\) corresponding to four different kind of nearest neighbour pairs with distances given by $d_1$, $d_2$, $d_3$ and $d_4$ as shown in Fig. 3(a) and Table I of the main manuscript. We used five different magnetic configurations as illustrated in Fig. S3. The values of exchange parameters calculated along with their corresponding nearest neighbour pair distances are given in Table S1.

\begin{table}[!h]
\centering
\renewcommand{\arraystretch}{1.5}
\label{tab1}
\caption{Magnetic Exchange Parameters Values}
\scalebox{1.1}{
\begin{tabular}{|c|c|c|}
\hline
\textbf{Pair Index(i)} & \textbf{Distance(d$_i$)(\AA)} & \textbf{Exchange parameter(J$_i$)(meV)}\\
\hline
1 & 4.14 & 2.01\\
\hline
2 & 4.65 & 0.11\\
\hline
3 & 4.18 & 0.59\\
\hline
4 & 4.56 & 0.01\\
\hline
\end{tabular}}
\end{table}